\documentclass[%
 reprint,
 amsmath,amssymb,
pre,
floatfix,
]{revtex4-1}

\usepackage{graphicx}
\usepackage{dcolumn}
\usepackage{bm}

\begin{document}


\title{The Partial Visibility Curve of the Feigenbaum Cascade to Chaos}

\author{Juan Carlos Nu\~no}\email{juancarlos.nuno@upm.es}
 \affiliation{Department of Applied Mathematics. Universidad Polit\'ecnica de Madrid, 28040 Madrid, Spain}

\author{Francisco J. Mu\~noz}
\affiliation{}%


\date{\today}

\begin{abstract}

A family of classical mathematical problems considers the visibility properties of geometric figures in the plane, e.g. curves or polygons. In particular, the {\it domination problem} tries to find the minimum number of points that are able to dominate the whole set, the so called, {\it domination number}. Alternatively, other problems try to determine the subsets of points with a given cardinality, that maximize the basin of domination, the {\it partial dominating set}. Since a discrete time series can be viewed as an ordered set of points in the plane, the dominating number and the partial dominating set can be used to obtain additional information about the visibility properties of the series; in particular, the total visibility number and the partial visibility set. In this paper, we apply these two concepts to study times series that are generated from the logistic map. More specifically, we focus this work on the description of the Feigenbaum cascade to the onset of chaos. We show that the whole cascade has the same total visibility number, $v_T=1/4$. However, a different distribution of the partial visibility sets and the corresponding partial visibility curves can be obtained inside both periodic and chaotic regimes. We prove that the partial visibility curve at the Feigenbaum accumulation point $r_{\infty} \approx 3.5699$ is the limit curve of the partial visibility curves ($n+1$-polygonals) that correspond to the periods $T=2^{n}$ for $n=1,2,\ldots$. We analytically calculate the length of these $n+1$-polygonals and, as a limit, we obtain the length of the partial visibility curve at the onset of chaos, $L_{\infty} =  L(r_{\infty}) \approx 1.0414387863$. Finally, we compare these results with those obtained from the period 3-cascade, and with the partial visibility curve of the chaotic series at the crossing point $r_c \approx  3.679$. 


\end{abstract}

\maketitle


\section{\label{sec:level1}Introduction}

Time series are being continuously generated from the multiple dynamic processes that take place in Nature: from the biggest scales of astrophysics, to the tiny scales of the genetic activity of genome, passing through the scales of many biological and sociological phenomena. Besides, new technological devices are enabling the gathering of larger amounts of information, that motivate the development of new and more powerful methodologies of analysis. Therefore, the study of time series has become one of the most relevant disciplines within the field of data science \cite{Cady}. These experimental signals are, in practice, of variable size, increasing in time as new data is being added but, necessarily of finite size. Precisely, the dependence of the properties of time series with size is one of the main aspects that must be carefully considered. It turns out that only in the case of rather large time series, certain intrinsic characteristics are clearly displayed, for instance high periodicities or, in the limit, chaos. However, the analysis of finite size time series have managed to show statistically conclusive results that, in most cases, allows a meaningful classification of this kind of signals \cite{Bradley}.

One of the recently proposed methods is based on the natural concept of visibility \cite{Gosh} that considers the interpolation of the discrete points $(t_i,x_i)$ for $i=1,2,\ldots,N$ of the time series as a contour in the plane. Accordingly, visibility defines a relationship between the points of the time series. Based on this concept, Lacasa {\it et al.} defined a map between a time series and a graph (network), the visibility map, in such a way that if two points of the time series can see each other then, their corresponding nodes in the visibility graph are (undirectedly) linked \cite{Lacasa1}. The relevance of this mapping is that, once the visibility graph is generated, the powerful tools of complex network theory can be applied to extract information about the properties of the time series. The success of this approach, either in its natural or horizontal version, has been remarkable. It has been applied to a large variety of problems during the last decade, in particular, to obtain unknown properties of chaotic, random or Brownian time series and to study spatial profiles that evolve over time in irreversible processes \cite{Luque3, Lacasa2,Luque1,Luque2, Bru}. All of these applications are only confirming the strength of visibility as a conceptual framework.


\begin{figure}[htpb]
\centering
\includegraphics[width=1.0\columnwidth]{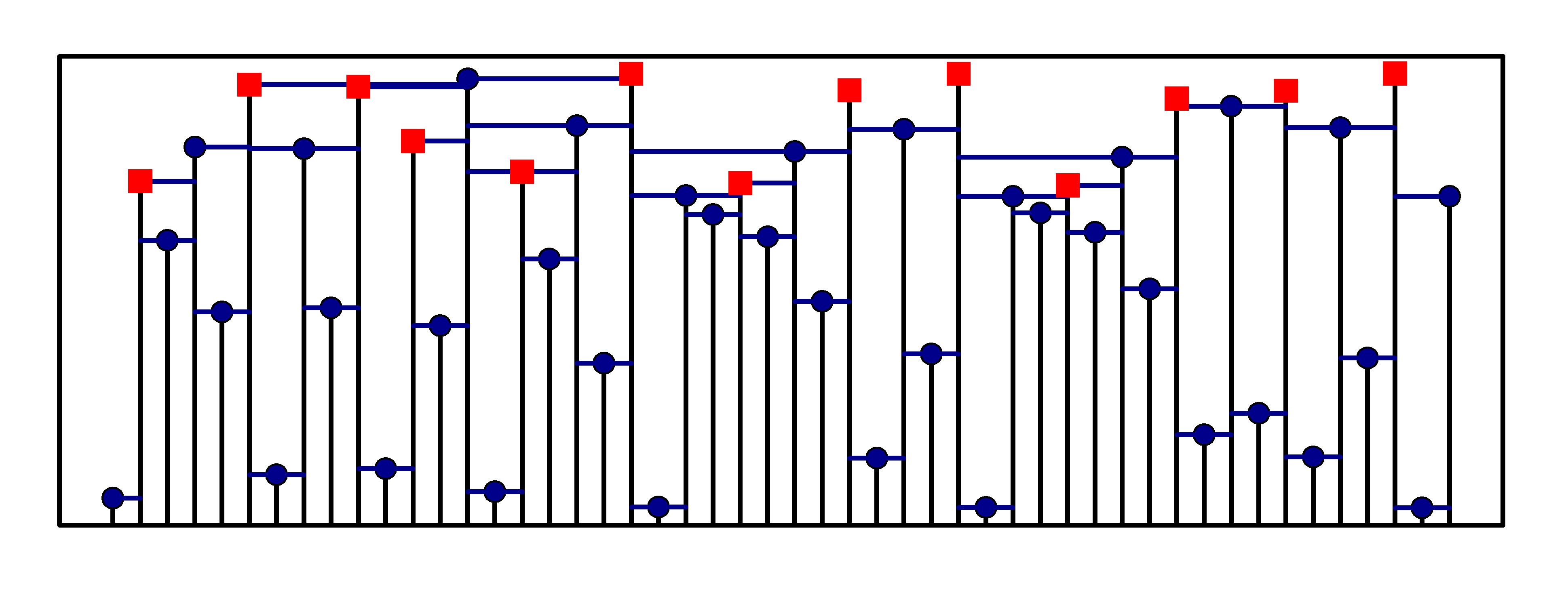}
\caption{Total visibility set resulting from the horizontal visibility map. Red squared points belong to the total visibility set and the blue circles are in their basin of visibility. The whole series (50 points) is seen from the 13 red points (including themselves). The total visibility is $13/50 \approx 1/4$.}
\label{vdom}
\end{figure}

Within this framework, an alternative approach considers visibility as a measure of {\it domination} \cite{Haynes}. Given a set of points in the plane that form a curve by pieces, for example, the points $(t_i,x_i)$ of a time series, it is necessary to identify the minimum set of points (lowest t cardinal) that dominates a maximum set of points of the whole series. In our context, domination means visibility. Therefore, the question is whether it is possible to find the set of minimum cardinality of the time series with a maximum basin of visibility. Classical problems that belong to this framework are the Eight Queens problem, the Art Gallery problem and Watch Tower problem \cite{Bell,Rourke,Cole}. Although the statement of these problems seems to be essentially geometrical, in  practice, finding their solution requires an enormous computational effort since, as already proven, this kind of optimization problems are NP-complete \cite{Garey}. 

There are several methods to compute the visibility sets \cite{Hedar,King}. Recently, genetic algorithms have been proposed as one of the most efficient and less computationally expensive methods \cite{Kubat, Alharbi}. We have used this type of algorithms to study the visibility properties of both periodic and chaotic series generated from the logistic equation \cite{Peitgen, Strogatz}. More specifically, we have determined two types of visibility sets: (i) the total visibility sets, i.e. minimum set of points that see the whole time series or, in other words, whose basin of visibility contains all of points of the time series, and (ii) the partial visibility sets of a given cardinality $m$, that are sets of exactly $m$ points which have a maximum basin of visibility, i.e. the basin of visibility set has a biggest cardinality.

The properties of the chaotic series and, in particular, the way they are generated from discrete dynamical systems are still questions of great relevance. This paper is mainly focused on determining the total and partial visibility properties of the time series that appear in the Feigenbaum cascade and the onset of chaos in the logistic model. Furthermore, in the search for universal explanations, we compare these results with the period 3 - cascade and with other chaotic series in the bifurcation diagram. After presenting the basic definitions of the model, the third section determines the total and partial visibility curves for the time series of the Feigenbaum cascade and its dependence with size. This section finalizes with a comparison with other chaotic regimes. The last section presents some conclusions we extracted from these results. Finally, an appendix details the genetic algorithm we have used to compute the visibility sets.

\subsection{\label{sec:level2}Definitions} 

A time series can be generated as a solution of an Intial Value Problem associated to the discrete difference equation:

\begin{equation}
x_{n} = f(x_{n-1}; r)
\end{equation}
for $n = 1,2, \ldots$ with the initial condition $x_0$. In general, it is assumed that $f$ is a continuous function of the real variable $x$ and the real parameter $r$. From the initial point $x_0$, this map generates an infinite sequence of real numbers: 
$\{x_k\}_{k \in \mathbb{N}}$. When $n$ represents time, this infinite sequence is usually referred as time series. 

\begin{figure}
\centering
\includegraphics[width=1\columnwidth]{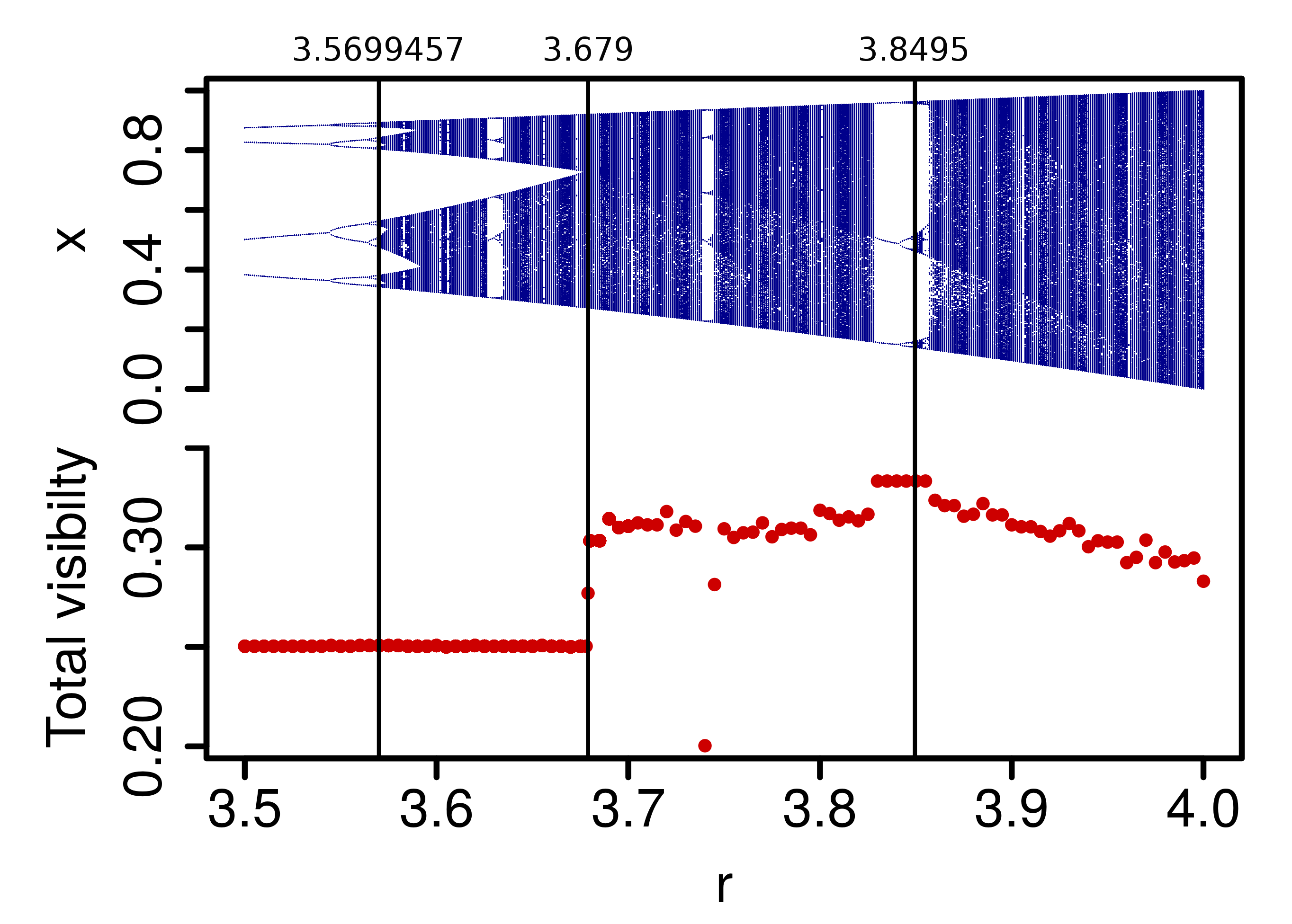}
\caption{Bifurcation diagram of the logistic map (top) together with the total visibility (bottom) as a function of the growth rate $r$. Vertical lines mark some significative $r$-values: the accumulation points of the two period dubling cascades corresponding to 2 and 3,$r(\infty) =3.5699457$ and $r_3(\infty) \approx 3.8495$, and the crossing point, $r_c \approx 3.679$, where main chaotic bands merge. As it can be seen, the total visibility is constant and equal to 0.25 for the Feigenbaum cascade and after the onset of chaos at $r(\infty)$. An abrupt change occurs precisely at the crossing point $r_c$. The same occurs in the 3-cascade: the total visibility is equal to 1/3 beyond the accumulation point $r_3(\infty)$ where its value starts to decrease}
\label{Rplot_2000_800_s500_in0p01_point0p5}
\end{figure}

One of the well known maps, inspired by the Verlhust population model, is the quadratic or logistic map:
\begin{equation}
f(x_n,r) = r \, x_n \, (1 - x_n)
\end{equation}
where $r$ is a growth rate. Despite its simplicity, this map gives rises to very complicated dynamics, as Robert May explained in his pioneer paper \cite{May}. As the growth rate $r$ increases, the behavior of the logistic map suffers multiple transitions. In particular, it includes the so called Feigenbaum cascade, that starts at $r= 1$ and reaches the onset of chaos at the accumulation point $r_{\infty} \approx 3.569945$. All these transitions are reflected in the bifurcation diagram in terms of the parameter $r$ (see the inset in figure \ref{Rplot_2000_800_s500_in0p01_point0p5}). Precisely, with regards to the study of chaotic dynamics, this map has provided an interesting example to verify many of the universal properties of chaos. It might seem surprising that after more than half a century studying this map, new properties continue to appear, helping to understand this type of dynamics.  For instance, the application of the visibility graph reveals some new characteristics of the bifurcation diagram \cite{Luque1,Luque2}.

In practice, only a finite number of points of the time series are observed. In many cases, this is not a problem since the behavior of the series can be predicted from the properties of the map $f$. For instance, if a unique stable fixed point for $f$ exists then, the time series tends to this point from almost all initial conditions. The same occurs when fixed points exist for successive powers of $f$ (periodic behavior). However, the exact prediction of the complete series could be practically unattainable in other situations where only partial information extracted from its finite size is available. This is the case of chaos, whose characterization requires in many cases the recourse to probabilistic techniques \cite{Schuster, Peitgen, Strogatz}.

The time series is an ordered set as induced by the time step, which differs from any set of points  that are generated from the map. This difference is crucial in what regards the visibility properties. In what follows, we present some definitions that will be used in the next sections.

{\it Horizontal visibility}.  Two points of the time series  $(t_i,x_i)$  and $(t_j, x_j)$ such that $t_i < t_j$ see horizontally each other if $x_k  < min\{x_i,x_j\}$ for all $t_k \in (t_i,t_j)$ (see figure \ref{vdom}).

{\it Basin of visibility of a point}. The set of points seen from a point $P_k(k,x_k)$ of the time series contour forms its basin of visibility, $B_k$. Formally,
\[
B_k=\{P_j\}_{j=1,2,\ldots,N}
\]

{\it Visibility of a point}. The cardinal of the basin of visibility of a point $P_k(k,x_k)$ is its visibility, $v_k$. That is:
\[
v_k = \#B_k
\]

{\it Basin of visibility of a set of points}. Given a set of points of the time series, $I_m$, the basin of visibility of this set is the union of the basin of visibility of all of the points that form $I_m$.
\[
BS_m = \bigcup_k \{B_k\}_{k \in I_m}
\]
with $m=1,2,\ldots,N$.

{\it Visibility of a set} It is the cardinal of the basin of visibility of the set $I_m$
\[
vS_m = \#BS_m
\]

{\it Visibility distribution of the time series}. The distribution function of the visibility of all of the points that form the times series is its visibility distribution. For each visibility number, $v$,  $P(v)$ represents the number of points of the time series that posses this visibility.

{\it Maximum visibility of one point}. Given a time series, the maximum visibility of one point is the largest visibility of any point of the time series:
\[
1V = max \{v_k\}_{k=1,2,\ldots,N}
\]
In other words, it is the maximum cardinality of the basin of visibility of any point.

{\it Partial visibility of $m$-points}. It is the maximum visibility of a set of $m$-points of the time series i.e. the largest cardinality of a basin of visibility of a set of points:
\[
mV = max \{ vS_m\}_{m =1,2, \ldots,N}
\]

These definitions refer to the number of points of the time series. Similar definitions can be used for the fraction of points that form each of the corresponding sets. For instance, 

{\it Partial visibility of a fraction $\mu$ of points of the whole series}. It is the maximum fraction of points that is seen from a set formed of a fraction $\mu$ of points of the whole series ($\frac{1}{N} \leq \mu \leq 1$).

Note that these fractions are always of the form: $pV_{\mu} = \frac{mV}{N}$ and, therefore, it verifies $\frac{1}{N} \leq pV_{\mu} \leq 1$.  The total visibility number corresponds precisely to $V= pV_1$.

It turns out that the maximum visibility of a set of $m$ points is the whole time series, that is $vS_m = N$. In this case, this means that this set has global visibility. Furthermore, the cardinal of this minimum set that sees all of the points of the series is called the total visibility.

{\it Total visibility}) It is the cardinal of the minimum subset of points that sees the whole series. In other words, it is the minimum set of points needed to cover the visibility of the whole time series from the union of their basins.

To carry out a numerical characterization of the partial visibility for each value of the parameter $r$ is not always an easy task.  It turns out that the size of the time series limits the period that can be identified. For obvious reasons, periodicities larger than the size of the series cannot be justified. Hence, for finite series, in practice, chaos cannot be differentiated from any large periodicity.  Even for lower periods, the computation of the partial visibility is affected at the bifurcation points.

\section{\label{sec4:level1}Visibility}

We compute the partial visibility properties of the time series from the Feigenbaum cascade to the onset of chaos. For each $r$-value, we obtain the $m$-visibility sets for $m=1, \ldots,m_{max}$, where $m_{max}$ is the cardinal of the total visibility set. To compare time series with different sizes, it is convenient to scale each of their visibility sets to the size of the series and describe them as a fraction of the whole series. Consequently, the parameter $\mu$ is in the interval $(0,1]$. Also, the fraction of the whole series that forms the visibility basin of the visibility set is denoted by $\nu(\mu)$ and its value belongs to $[0,1]$.

The calculation of the distribution of the basin of visibility of the time series for any periodic regimes can be obtained based on its period, which doubles as the parameter $r$ varies. If the periods of the cascade are written as $T = 2^k$, their visibility distributions (including the points that see their own basins) are those described in table \ref{tab1}. For the case $k=1$, this means that half of the points of the time series have a basin of visibility of cardinal 3 and the other half have a cardinal 5 (including the points that see their own basins). As the period increases, i.e. $k$ increases,  new partitions of the whole series appear, with subsets of less cardinality but which are formed by points with a larger basin of visibility. In general, for period $T=2^k$, the series is divided into $k+1$ sets with an increasing basin of visibility. The largest class has cardinality $2^{-k}$ and it is formed by points that see $2 \, k+3$ points, including themselves.

\begin{table}[tbp]
\caption{Visibility distributions of the period doubling cascade $T = 2^k$. For each $k$, the points of the series are divided into classes according to their visibility. For instance, for $k=1$, the series is formed by two classes: half of points have a basin of visibility of 3 points (including itself) and the other half see 5 points (including itself). As $k$ increases the fraction of points with maximum visibility tends to 0 exponentially. These distributions are reflected in the form of the visibility polygonals for each period as depictd in figures 3 and 4.} 
\label{tab1}
 \small
\begin{tabular}{cccccccccc}
\hline \hline

\textbf{k} & \textbf{3}      & \textbf{5}     & \textbf{7}     & \textbf{9}      & \textbf{11}     & ... & \textbf{2k+1}      & \textbf{2k+3}    & ...  \\
\hline
\textbf{0} & 1   &     &     &      &      &   &           &        &   \\

\textbf{1} & 1/2 & 1/2 &     &      &      &   &           &       &    \\

\textbf{2} & 1/2 & 1/4 & 1/4 &      &      &   &           &       &    \\

\textbf{3} & 1/2 & 1/4 & 1/8 & 1/8  &      &   &           &       &    \\

\textbf{4} & 1/2 & 1/4 & 1/8 & 1/16 & 1/16 &   &           &      &     \\

...          &     &     &     &      &      &   &           &       &    \\

\textbf{k} & 1/2 & 1/4 & 1/8 & 1/16 & 1/32 & ... & $1/2^k$   & $1/2^k$ & ... \\
\hline \hline
\end{tabular}
\end{table}

For obvious reasons, to detect periodicities for large values of $k$ requires a sample of series of equivalent sizes. Besides, as the period increases, the number of points with maximum visibility $2 \, k + 3$ decreases at a larger rate $2^{-k}$. At the limit, for a infinite time series, the probability of finding a point with the maximum visibility tends to 0.

\begin{figure}
\centering
\includegraphics[width=1.0\columnwidth]{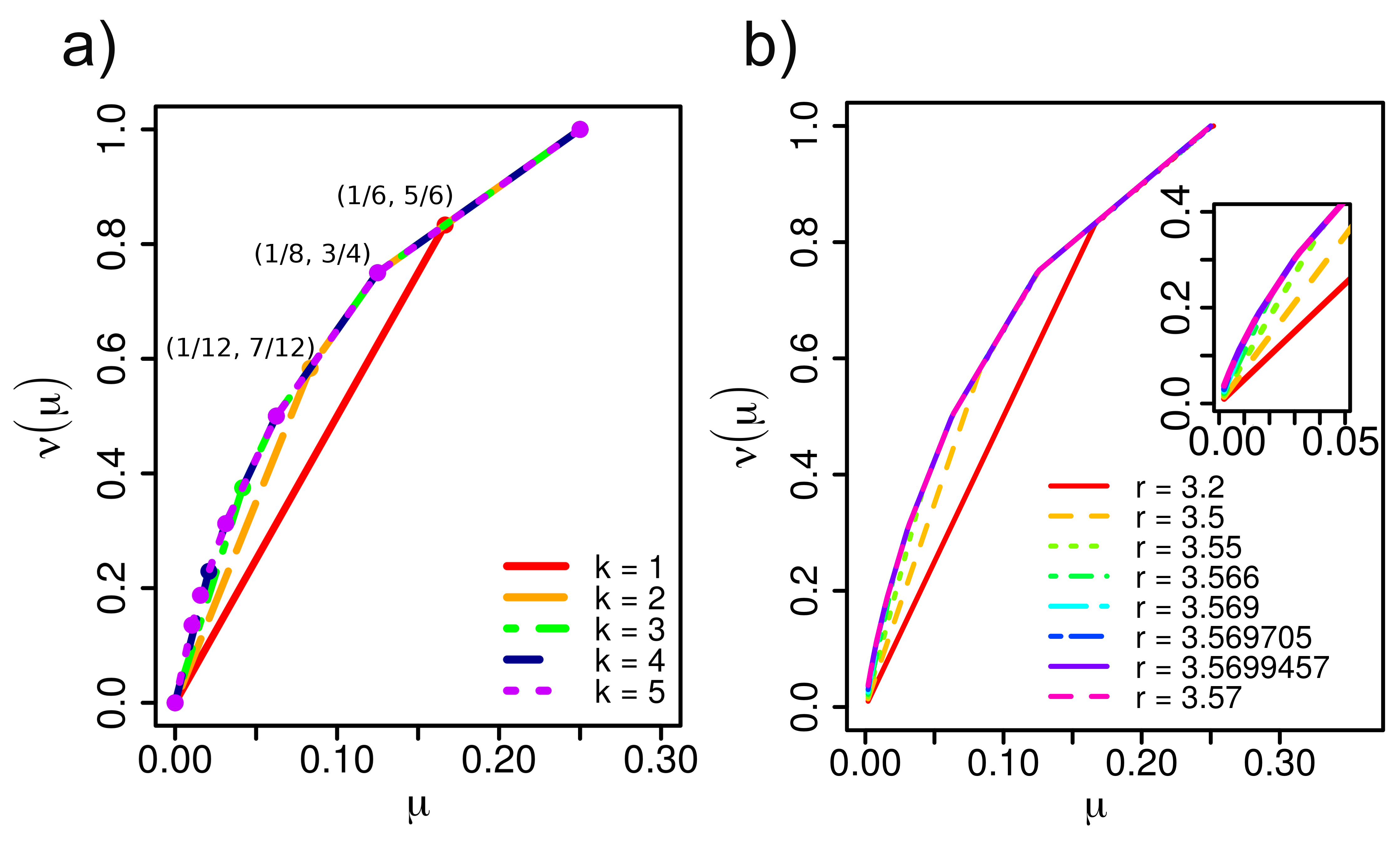}
\caption{(a) The partial visibility curves can be theoretically determined from visibility distribution of points of the time series for each period $T = 2^k$ for $k=1,2,\ldots$. This figure depicts the visibility curves for time series with the first five periods of the Fiegenbaum cascade. As it can be seen, the dependence of the fraction of the series that is seen from a given fraction of points is are polygonals with a number of segments that depends on the period. Concretely, for each $k$ the corresponding polygonal has $k+1$ segments. It is worthy to observe that the polygonal of a given period (with $k > 1$) contains part of the pieces of the polygonals of lower periods. (b) Computational visibility curves for different $r$ values. The size of the series is 500 points.}
\label{Figpolyteor}
\end{figure}

This distribution of visibility of the points of the time series determines its partial visibility. In Figure \ref{Figpolyteor} the visibility curves for the periods $k=1,2,3,4$ are depicted. As it can be seen, these polygonals are formed by $k+1$ segments of decreasing slope. These polygonal curves appear for every periodic series of period $T=2^k$ and they can be described mathematically as follows. For period $T=2$ ($k=1$):

\begin{equation}
\nu_1(\mu) = 
\begin{cases}
5 \, \mu  & 0 \leq \mu \leq \frac{1}{6} \\
2 \, \mu + \frac{1}{2}  & \frac{1}{6} \leq x \leq \frac{1}{4}
\end{cases}
\end{equation}
For period $T=4$ ($k=2$), an additional segment appear for small values of $\mu$:
\begin{equation}
\nu_2(\mu) = 
\begin{cases}
7 \, \mu  & 0 \leq \mu \leq \frac{1}{12} \\
4 \, \mu + \frac{1}{4}  & \frac{1}{12} \leq \mu \leq \frac{1}{8} \\
2 \, \mu + \frac{1}{2}  & \frac{1}{8} \leq \mu \leq \frac{1}{4}
\end{cases}
\end{equation}
The third period doubling corresponds to $k=3$ and the partial visibility curve is:
\begin{equation}
\nu_3(\mu) = 
\begin{cases}
9 \, \mu  & 0 \leq \mu \leq \frac{1}{24} \\
6 \, \mu + \frac{1}{8}  & \frac{1}{24} \leq \mu \leq \frac{1}{16} \\
4 \, \mu + \frac{1}{4}  & \frac{1}{16} \leq \mu \leq \frac{1}{8} \\
2 \, \mu + \frac{1}{2}  & \frac{1}{8} \leq \mu \leq \frac{1}{4}
\end{cases}
\end{equation}

In general, the polygonal for the period $T=2^k$ is the graph of the function:

\begin{equation}
 \nu_k(\mu) = 
\begin{cases}
(2 \, k + 3) \, \mu  & 0 \leq \mu \leq \frac{1}{3} \frac{1}{2^k}  \\
2 \, k \, \mu + \frac{1}{2^k} &  \frac{1}{3} \frac{1}{2^k} \leq \mu \leq \frac{1}{2^{k+1}} \\
2 \, (k-1) \, \mu + \frac{1}{2^{k-1}} &  \frac{1}{2^{k+1}} \leq \mu \leq \frac{1}{2^{k}}  \\
... & ... \\
4 \, \mu + \frac{1}{4}  & \frac{1}{16} \leq \mu \leq \frac{1}{8} \\
2 \, \mu + \frac{1}{2}  & \frac{1}{8} \leq \mu \leq \frac{1}{4}
\end{cases}
\end{equation}

\begin{figure*} 
\centering
\includegraphics[width=\textwidth]{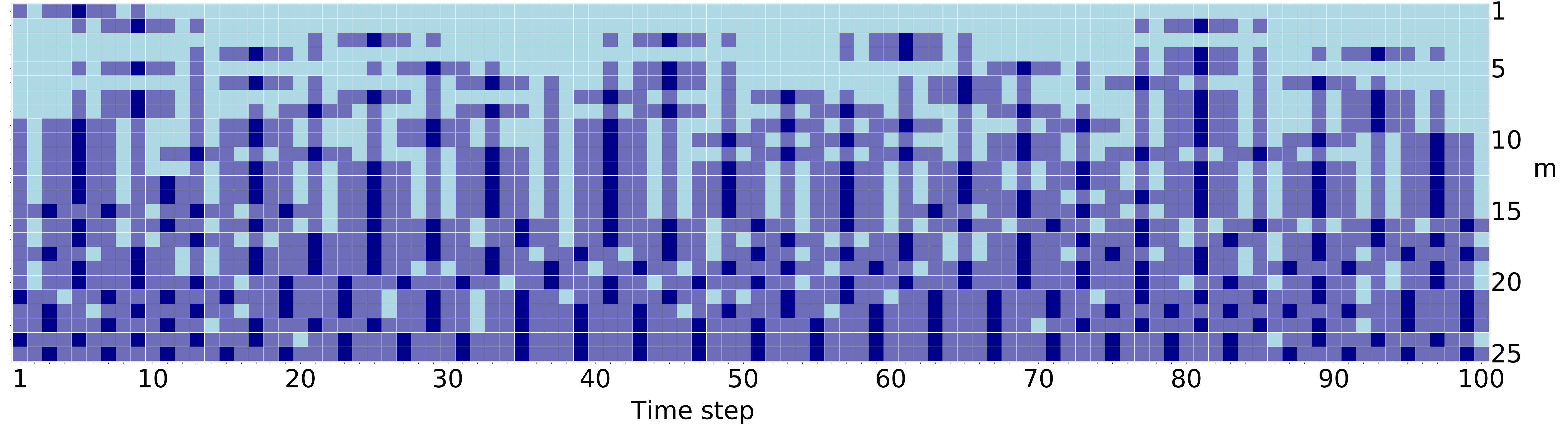}
\caption{Schematic representation of how the $m$-visibility sets are formed for each value of $m=1,2,ldots,25$ (vertical axis) in a example  of time series of size 100 points (horizontal axis).  The top row  depicts the point selected to form the 1-visibility set (darkest blue) and its basin of visibility (dark blue). The rest of the points (lighter blue) of the sample, i.e. 93 points (columns)  are not seen from this  1-visibility set. The second row corresponds to $m=2$: an additional  point with not overlapping basin of visibility with the previously  1-visibility set is chosen. In this way, this 2-visibiity set has doubled the cardinality of the basin of visibility with respect to $m=1$. It turns out that for a certain $m$-value, in this example for $m=9$, adding a new point to the visibility set can not be done without overlapping its basin of visibility with that of the other points. At this moment, the slope of the segment in the polygonal curve decreases (see figure cite{Figpolyteor}). The time series has been generated with the value of the growth rate:  $r=3.5$.}
\label{plotgg}
\end{figure*}

It is not difficult to visualize the appearance of these polygonals (see Figure \ref{plotgg}). In order to form the partial visibility set of a single point, we choose one of the points with the largest basin of visibility (darkest pixel in the first row). Next, to form the visibility sets of two points, we choose two points with the maximum basin of visibility and, in order to have the largest union of their basins of visibility, we take these two points with non-overlapping basins. The visibility sets of increasing cardinality are formed by adding points that result in the largest basin of visibility. Whereas the basins of visibility are not overlapping with the rest of points of the visibility set, the addition of a new one increases in the same amount the partial visibility, i.e. the union of their basins of visibility. It turns out that, there is a step such that the addition of a new point to the visibility set cannot be done without overlapping with the rest of the basins of other points of the visibility sets. At this moment, consequently, the partial visibility of this set does not increase in the same amount as the visibility sets of lower cardinality. At this step, a new segment begins with a lower slope. As it can be observed in figure \ref{Figpolyteor}, in the case of period $T=2$, this occurs at $\mu=1/6$ ($\nu(1/6) = 5/6$), whereas in the case of $T=4$, the first vertex of the polygonal occurs at $\mu=1/12$ ($\nu(1/12) = 7/12$) and the second at $\mu =1/8$ ($\nu(1/8) = 3/4$).

The number of segments of each polygonal, $s$, increases logarithmically with the period, $T=2^k$ as $s = \log_2(T) + 1 = k+1$, for $k=1,2,\ldots$. Furthermore, the $k$-polygonal contains part of the pieces of the $k-1$ polygonal with an additional segment.  The length of the added segments decreases with the period and tends to zero.
At the limit, when $k \to \infty$ and the period $T$ tends to infinity, the partial visibility curve has an infinite number of segments.  This limit curve, that contains all of the polygonals of the periodic series in the cascade, is the partial visibility curve at the onset of chaos that take place at the accumulation point $r_{\infty}$.

The partial visibility curves of the periodic series are independent of the size of the series. In practice, we can only obtain polygonals from finite size samples. Since infinite periods only occurs in infinite series, the computation of the partial visibility polygonals at the limit of infinite periods (chaos) is only approximate and, therefore, could depend on the size of the series. This dependence is specially relevant for the slope at the origin because, as the period increases new segments appear for values of $\mu$ approaching 0. For periodic series, consequently, the slope, $a$, of the first segment starting at the origin is independent of the size and depends on the period $T = 2^k$ as $ a = 2 \, \log_2 (T) +3 =  2 \, k + 3$. 

On the contrary, for chaotic series the 1-partial visibility of one point, i.e. the value of the right limit of the partial visibility curve at the origin, depends on the size of the time series, $N$.  As it can be seen, the 1-partial visibility increases as a power law with the size $N$ and can be approximately described by:

\begin{equation}
v_1(N) = 2.186 N^{-0.845}
\end{equation}

with an statistical $R^2 = 0.999$. The exponent of the power law is characteristic of the chaotic signal and, consequently, depends on the parameter $r$. Note that, for periodic time series, a power law also represents the fraction of the maximum visibility of a point versus the size of the series but, in these cases, the exponent is $-1$.



The other extreme of the polygonals is the point $(\mu= 1/4,\nu=1)$ for all the $r$-values in the Feigenbaum cascade. This indicates that the total visibility of the series, i.e. the minimum set of points that sees the rest of the time series is independent of $r$. For the Feigenbaum cascade, it is only needed one quarter of the points of the series to see the whole series. Furthermore, it is interesting to remark that, as shown in figure \ref{Rplot_2000_800_s500_in0p01_point0p5}, this value of the total visibility remains unchanged beyond the accumulation point $r_{\infty}$ until the crossing point at $r_c = 3.67...$ \cite{Livadiotis}, where it changes to approximately one third. Besides, for all of this range, this value of the total visibility is independent of the size of the time series. 



\begin{figure} 
\centering
\includegraphics[width=1.0\columnwidth]{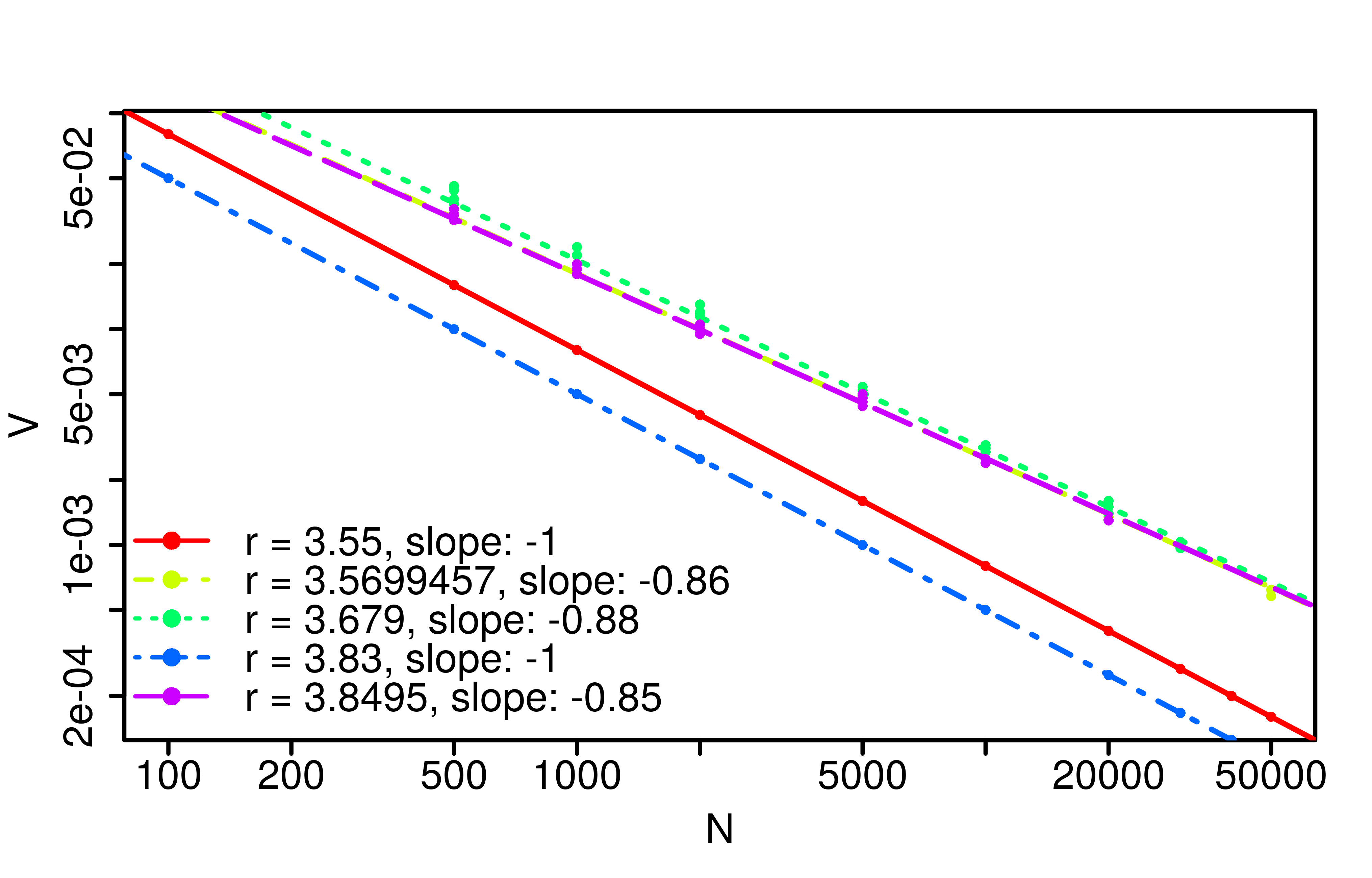}
\caption{Contrary to the periodic time series, the maximum visibility of one point of the series, i.e. the slope of the partial visibility curve at the origin depends on the size of the series.}
\label{pwlaw}
\end{figure}


\subsection{\label{sec41:level2}Length of the partial visibility curves}

As explained in the previous section, there is a partial visibility curve for each time series in the Feigenbaum cascade to the onset of chaos. The limit curve of these polygonals, which has an infinite number of segments of decreasing length, corresponds to the onset of chaos (infinite period) at the accumulation point $r_{\infty}$. The question is whether the properties of the partial visibility curve are representative of the period doubling to the onset of chaos. 

First at all, we compute the length of each of the polygonals and obtain the length of the limit curve that, as mentioned above, corresponds to the chaotic dynamics at the accumulation point $r_{\infty}$. The length of the partial visibility $k$-polygonal for each period $T=2^k$ can be computed as the sum of the length of the segments that form the curve:
\[
L_k = \sum_j^s l_{jk}
\]
where $l_{jk}$ is easily calculated by using the Pythagorean theorem and $s$ is the number of segments of each polygonal. Since the $k$-polygonal contains part of the $k-1$-polygonal, the $k$-length $L_k$ can be computed iteratively:

\begin{widetext}
\begin{eqnarray}
L_k  & = & \sum_{i=1}^{k-1}  \sqrt{\left(\frac{1}{2^{i+1}}-\frac{1}{2^{i+2}} \right)^2 + \left( 2 \, i  \, \left(\frac{1}{2^{i+1}} \right) + \frac{1}{2^i} - ( 2 \, i + 2 ) \, \left( \frac{1}{2^{i+2}} \right) -  \frac{1}{2^{i+1}} \right)^2} + \\ \nonumber
& +  & \sqrt{\left( \frac{1}{2^{k+1}} - 
\frac{1}{3 \, 2^k} \right)^2  + \left(2 \, k \, \left( \frac{1}{2^{k+1}}  \right)  + \frac{1}{2^k} - \left(2 \, k \frac{1}{3 \, 2^k} + \frac{1}{2^k} \right) \right)^2} + \\ \nonumber
& + &   \sqrt{\left(\frac{1}{3 \, 2^k} \right)^2 + \left(2 \, k \, \left(\frac{1}{3 \, 2^k} \right) + \frac{1}{2^k} \right)^2}
\end{eqnarray}
\end{widetext}

From this point, the limit for $k$ to $\infty$ can be computed and yields the length of the limit curve at the onset of chaos $r=r_{\infty}$.

\begin{equation}
L_{\infty} \approx  1.0414387863956818
\end{equation}
with an accuracy larger than $10^{-16}$.  

Information about how this limit is attained can be obtained from the ratio between successive polygonal length differences. More specifically:

\[
\phi_n = \frac{L_n-L_{n-1}}{L_{n+1}-L_n} 
\]

It can be straightforwardly proven that 

\[
\lim_{n \to \infty} \phi_n = 2
\]
Most probably, this value is a reminiscent of the period doubling that occurs in the cascade to the onset of chaos at $r=r_{\infty}$. 

Following similar steps as in the Feigenbaum cascade, it is possible to find the visibility distribution of the time series for the period three cascade. Concretely, this distribution is represented in \ref{tab2} for each of the periods $T=3 \, 2^{k-1}$ for $k =1, 2, \ldots$

\begin{table}[tbp]
\caption{Visibility distributions for the time series generated in the doubling period cascade $T=3 \, 2^{k-1}$. Similar as occurs in the Feigenbaum cascade, as the period increases the maximum visibility of one point increases. For $k=1$ the whole series is equally divided into points with a basin of visibility formed by 3, 4 and 6 points. For a given $k$, a fraction $1/3 \, 2^{-k+1}$ of the series have the maximum vibility per point $2 \, (k+2)$.}  
\label{tab2}
 \small
\begin{tabular}{cccccccccc}
\hline \hline
\textbf{k} & \textbf{3}      & \textbf{4}  & \textbf{6} & \textbf{8}    & \textbf{10}   & ... & \textbf{2(k+1)}   & \textbf{2(k+2)}  & \ldots    \\
\hline
\textbf{1} & 1/3 & 1/3 & 1/3 &      &       &   &                      &    &                    \\

\textbf{2} & 1/3 & 1/3 & 1/6 & 1/6  &       &   &                      &            &           \\

\textbf{3} & 1/3 & 1/3 & 1/6 & 1/12 & 1/12  &   &                      &          &             \\

\textbf{4} & 1/3 & 1/3 & 1/6 & 1/12 & 1/24  &   &                      &       &                \\

\ldots          &     &     &     &      &       &   &                      &            &           \\

\textbf{k} & 1/3 & 1/3 & 1/6 & 1/12 & 1/24  &  ... & $1/(3\cdot 2^{k-1})$   & $1/(3\cdot2^{k-1})$   &  \ldots  \\
\hline \hline
\end{tabular}
\end{table}

This visibility distribution of the points of the time series induces the following family of $(k+2)$-polygonals for the period three cascade. For period $T=3$ ($k=1$):

\begin{equation}
\nu_1(\mu) = 
\begin{cases}
6 \, \mu  & 0 \leq \mu \leq \frac{1}{9} \\
3 \, \mu + \frac{1}{3} & \frac{1}{9} \leq \mu \leq \frac{1}{6} \\
\mu + \frac{2}{3}  & \frac{1}{6} \leq \mu \leq \frac{1}{3}
\end{cases}
\end{equation}
and for period $T=6$ ($k=2$):
\begin{equation}
\nu_2(\mu) = 
\begin{cases}
8 \, \mu  & 0 \leq \mu \leq \frac{1}{18} \\
5 \, \mu + \frac{1}{6} & \frac{1}{18} \leq \mu \leq \frac{1}{12} \\
3 \, \mu + \frac{1}{3} & \frac{1}{12} \leq \mu \leq \frac{1}{6} \\
\mu + \frac{2}{3}  & \frac{1}{6} \leq \mu \leq \frac{1}{3}
\end{cases}
\end{equation}
The polygonal for a general period $k$ has $k+2$ segments and it can be described as:
\begin{equation}
 \nu_k(\mu) = 
\begin{cases}
(2 \, k + 4) \, \mu  & 0 \leq \mu \leq \frac{1}{9} \frac{1}{2^{k-1}}  \\
(2 \, k +1) \, \mu + \frac{1}{3 \, 2^{k-1}} &  \frac{1}{9} \frac{1}{2^{k-1}} \leq \mu \leq \frac{1}{3} \frac{1}{2^{k}} \\
(2 \, k - 1 ) \, \mu + \frac{1}{3 \, 2^{k-2}}  &  \frac{1}{3} \frac{1}{2^{k}} \leq \mu \leq \frac{1}{3} \frac{1}{2^{k-1}}  \\
... & ... \\
3 \, \mu + \frac{1}{3}  & \frac{1}{12} \leq \mu \leq \frac{1}{6} \\
 \mu + \frac{2}{3}  & \frac{1}{6} \leq \mu \leq \frac{1}{3}
\end{cases}
\end{equation}

Based on this general equation, it is easy to find the length of each $(k+2)$-polygonals that corresponds to period $T=3 \, 2^{k-1}$ for $k=1,2, \ldots$, as we have done for the Feigenbaum cascade:

\begin{widetext}
\begin{eqnarray}
L_k  & = & \sum_{i=1}^{k}  \sqrt{\left(\frac{1}{3 \, 2^{i-1}}-\frac{1}{3 \, 2^{i}} \right)^2 + \left( (2 \, i - 1) \, \left(\frac{1}{3 \, 2^{i-1}} \right) + \frac{1}{3 \, 2^{i-2}} -  ( 2 \, i + 1 ) \, \left( \frac{1}{3 \, 2^{i}} \right) -  \frac{1}{3 \, 2^{i-1}} \right)^2} + \nonumber \\ 
& +  & \sqrt{\left( \frac{1}{3 \, 2^{k}} - 
\frac{1}{3^2 \, 2^{k-1}} \right)^2  + \left( (2 \, k +1) \, \left( \frac{1}{3 \, 2^{k}} \right)  + \frac{1}{3 \, 2^{k-1}} - 2 \, (k + 2) \, \frac{1}{3^2 \, 2^{k-1}}  \right)^2} + \nonumber \\ 
& + &   \sqrt{\left(\frac{1}{3^2 \, 2^{k-1}} \right)^2 + \left(2 \, (k+2) \, \left(\frac{1}{3^2 \, 2^{k-1}} \right)  \right)^2}
\end{eqnarray}
\end{widetext}

The limit for $k \to \infty$ yields the length of the chaotic time series at the accumulation point $r_3(\infty) \approx 3.8495$:
\[
L_{\infty} =1.089156529597995
\]
with an error less than $10^{-16}$. As before, it can be proven that the ratio of the length increments, $\phi_n$, for this cascade also tends to 2 as $n \to \infty$, likely reflecting the period doubling generation of these partial visibility curves.  

The estimation of the length of chaotic series is, in general, a difficult task that depends on the size of the series. Contrary to what occurs with the polygonals that corresponds to the accumulation points of the period doubling cascades $r(\infty)$ and $r_3(\infty)$, whose length can be calculated analytically as a limit of lengths, the length of other chaotic polygonals can only be estimated numerically. As an example, we present the results obtained for the partial visibility curve at the crossing point at $r_c \approx 3.679$. This is the point where the total visibility changes with respect to the previous chaotic time series that occurs between $r(\infty)$ and this $r$-value.



\begin{widetext}

\begin{table}[tbp]
\caption{Lengths of the partial visibility curves for some $r$-values. We show the values obtained numerically for different sizes and compare with the theoretical ones, when possible. This can be done for chaotic time series in the two accumulation points of the period doubling cascades of period 2 and 3, specifically for $r (\infty) \approx 3.5699457$ and $r_3(\infty) = 3.8495$. On the contrary, we only have the numerical estimation for the chaotic time series at $r \approx r_c = 3.679$. Note that, as stated in the main text, the length of the visibility curves increases with the period doubling in both cascades} 
\label{tab3}
 \small
\begin{tabular}{cccccc}
\hline \hline

\textbf{r} & \textbf{P}   & \textbf{$L_{100}$}  & \textbf{$L_{200}$}     & \textbf{$L_{500}$}      & \textbf{$L_{theor}$}  \\
\hline
3.2        & 2            & 1.03700316930642   &  1.03603504624364   &  1.03918248184197   &   1.03617558372378             \\
3.5        & 4            & 1.04728686690324   &  1.04053195981291   &  1.04184640637487   &   1.04056021591033             \\
3.55       & 8            & 1.04747676315611   &  1.0412021148318    &  1.04189317986302   &   1.04123453225885             \\
3.566      & 16           & 1.04319618940498   &  1.04136824297737   &  1.04197872868395   &   1.04138246726324             \\
3.569      & 32           & 1.04113976319583   &  1.04140573758477   &  1.04160016531433   &   1.04142152414595             \\
3.569705   &  64          & 1.04427042122209   &  1.04141497645852   &  1.0418198242307    &   1.04143309632053             \\
3.5699457  & Chaotic      & 1.04218331587125   &  1.04143050731735   &  1.04206770916682   &   1.04143878639568           \\
3.679      & Chaotic      & 1.04427071684744   &  1.04487320574774   &  1.07049103765783   &                                \\
3.83       & 3            & 1.09361370780548   &  1.08117852726087   &  1.08510130862526   &   1.08724685599356           \\
3.845      & 6            & 1.09702929802809   &  1.08276135781289   &  1.0865197551801    &   1.08876803791482            \\
3.848      & 12           & 1.0867102615155    &  1.08916529939347   &  1.09582884912951   &   1.08905703200518            \\
3.8493     & 24           & 1.0871304606573    &  1.09122875720526   &  1.09425146906674   &   1.08912744605451            \\
3.8494     & 48           & 1.08711109502563   &  1.09132562939643   &  1.09173106419493   &   1.08914725694129            \\
3.8495     & Chaotic      & 1.09116430606687   &  1.08725619071464   &  1.09738454932656   &   1.08915652959799           \\ 

\hline \hline
\end{tabular}
\end{table}

\end{widetext}

\section{\label{sec42:level2}Conclusions}

A time series can be viewed as an ordered set of points in the plane and, therefore, it can be studied using a geometric approach in terms of its visibility properties. In particular : (i) the total visibility, which gives the minimum number of points that are needed to see the whole time series and provides information about the visual accessibility; (ii) the partial visibility, which computes the maximum visibility of a subset of $m$ points of the time series, for $1 \leq m \leq N$ and, in particular, when $m=1$, the maximum visibility of a point represents geometrically the slope at the origin of the curve of the partial visibility. Certainly, the computation of the partial visibility curve for each $m$ provides a complete set of information but, unfortunately, it would require a big computational effort that cannot be easily carried out over the whole range of variation of the growth rate $ r$, i.e. $1 \leq r \leq 4$.

The partial visibility of a given number of points $m$ or of a fraction of points $(\nu=m/N)$ depends on the period of the time series. The partial visibility of a single point ($m=1$) always increases with the period of the series and, in the limit tends to infinity as the size of the time series goes to infinite. On the contrary, for $m >1$ there is a period from which the value of the partial visibility stops increasing with $\nu$. In fact, for $ \nu = \nu_c = \frac{1}{4}$, the partial visibility covers the whole series and coincides with the total visibility for all periods and beyond the onset of chaos at $r_{\infty}$. For $\nu > \nu_c$ the addition of new points to the visibility set does not imply an increase in its visibility. Precisely, the total visibility value suddenly changes at the crossing point  $r_c \approx 3.679$ that coincides with the point where chaotic bands merge \cite{Livadiotis}. From a geometrical point of view, it could be said that the Feigenbaum cascade ends at this point $r_c$. A similar consideration can be done with regards to the cascade of period 3, where the abrupt change of the total visibility occurs beyond the accumulation point at $r \approx 0.86$.

It is not easy to compute the total or the partial visibility properties of a time series. We have opted for a genetic algorithm to solve this problem with, in our opinion, satisfactory results. Genetic algorithms apply an evolutionary and selective process to the genotypes that form the population according with a simple codification:  each digit 1 in the genotype codifies an element of the visibility set, i.e. a set of points that has a required basin of visibility; on the contrary, a digit 0 means that this point is not visible from any point of a visibility set.  A fitness function is assigned to each genotype that quantifies the visibility of the codified set according to the optimization problem to be solved, either (i) to find the sequences with the lowest number of 1 that see the whole set of points of the series (total visibility problem) or (ii) to find the genotype with a given a number of 1 in the sequence with the largest basin of visibility (partial visibility problem).

In our view, the analytical version of this partial visibility approach is an interesting mathematical challenge. Given a continuous curve, finding the minimum interval that sees either a fraction or the whole curve.The continuous formulation of the visibility problem would allow to apply powerful techniques of functional analysis and, most likely, would provide new analytical results.

\appendix

\section{\label{sec3:level1}Genetic algorithm for the computation of the visibility sets}

The determination of the visibility properties of a time series requires the computation of the minimum set of points that sees a given proportion of the points of the time series. This process is not evident and it is, in fact, very costly in computational terms \cite{Hedar,Rourke,King}. In this paper, we have applied a genetic algorithm as it is one of the best options known in the literature \cite{Alharbi}.

\begin{figure}
\centering
\includegraphics[width=0.8\columnwidth]{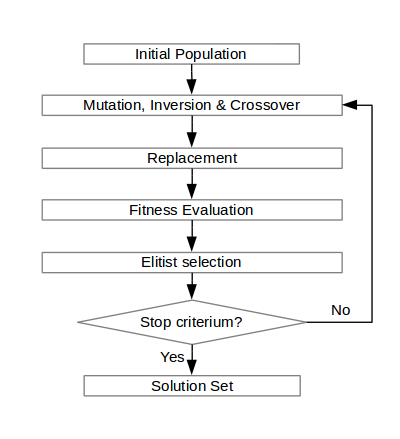}
\caption{Flowchart of the genetic algorithm.}
\label{Figflowchart}
\end{figure}

The flowchart of this genetic algorithm is depicted in figure \ref{Figflowchart}. The genetic algorithm simulates an evolutionary process over a population of $N$ genomes of size $\nu$ that equals the size of the time series (number of points) \cite{Kubat, Goldberg}. Each position of the genomes informs whether the corresponding point of the time series belongs to a visibility set. In particular, if at the position $i$ the value of the genome is 1, then the corresponding point of the time series $(t_i, x(t_i))$ belongs to the visibility set (partial or global). On the contrary, if its value is 0 then, the corresponding point does not belong to a visibility set.

We associate to each genome a fitness value that measures the range of visibility of the codified visibility set and the number of points that form the visibility set. The optimization process seeks the genotypes that sees the maximum set of points with the minimum cardinal of the visibility sets (minimum number of points).  Although the fundamentals of the genetic algorithm applied to solve both the total and the partial visibility problems are quite similar, their corresponding fitness functions differ. In the total visibility problem, by definition, the visibility set has to see the whole series, i.e. the basin of visibility must be the entire set of points of the time series. A simple way to carry this out requires defining the fitness of genome $k$ as:

\begin{equation}
f_k = 
\begin{cases}
\frac{SB_k}{vS_k} & if  \hspace*{5mm} SB_k = N \\
0 & if \hspace*{5mm}  SB_k \neq N
\end{cases}
\end{equation}
where $vS_k = \#1$ of the corresponding genome.

On the other hand, the partial visibility problem seeks to find the minimum visibility set that sees a given fraction of the time series. In this case, it is convenient to define a different fitness function:
\begin{equation}
f_k = 
\begin{cases}
\frac{SB_k}{vS_k} & if  \hspace*{5mm} vS_k = v \\
0 & if \hspace*{5mm}  vS_k \neq v
\end{cases}
\end{equation}
where $v$ is the value of the partial visibility problem we want to solve. For instance, for a 1-visibility set, $v=1$.

Note that, in both cases, we are searching for the maximum value of $f_k$, either by minimizing the denominator (total visibility) or by maximizing the numerator (partial visibility).

The main steps of the algorithm are:
\begin{enumerate}

\item Generate randomly an initial population of $N$ genomes formed by binary digits $\{0,1\}$. For global visibility, we take $N = \nu$ for all sequences. Whereas for partial visibility, the number of 1 in the genotype coincides with the cardinal of the visibility set we are looking for. For instance, for 1-visibility searching the initial population is formed by genomes with only one 1 and $\nu -1$ 0, randomly located.

\item At each step, a new population is generated  by replication from the previous population. Each individual of the population gives rise to another individual according to:

\begin{enumerate}
\item {\it Point Mutations} With a given probability $p$ any sequence of the population produces a mutant. This mutant contains a proportion of random modifications of the parent sequence. Both the mutation probability and the number of point mutations that occur are parameters that can be tuned to minimize the time of convergence. 

\item {\it Inversion} A number of individuals $n$ are chosen to invert part of their genome. A randomly selected sequence of the genome is inverted and placed in the original location. 

\item {\it Crossover} A number $c$ of pairs of individuals of the population share two random parts (of equal length) of their genomes (crossover) to create a new individual. The crossover length is also a parameter of the model and it is also tuned adequately to lower the computation time. We take this length randomly.

\end{enumerate}

The new population is added to the previous one, doubling as a consequence the number of individuals  

\item This doubled population, obtained by replication, is ordered according to its fitness value. Then, we classified the best $N$ genomes to survive to the next generation. 

\item The algorithm stops when either an error function attains a definite value or 
a fixed number of iterations is carried out. This error function is defined from a convergence criterion, for instance, when the average level of fitness of the population stops increasing.

\end{enumerate}

The program is written using R \cite{CRAN}. In all of the simulations we performed, the parameter setup was: 
\[
N = 100; \, p = 0.01; \, c = 10; \, n = 10
\]
In order to assure convergence, all simulations stop when the number of iterations is $10^4$.

\begin{figure}
\centering
\includegraphics[width=1.0\columnwidth]{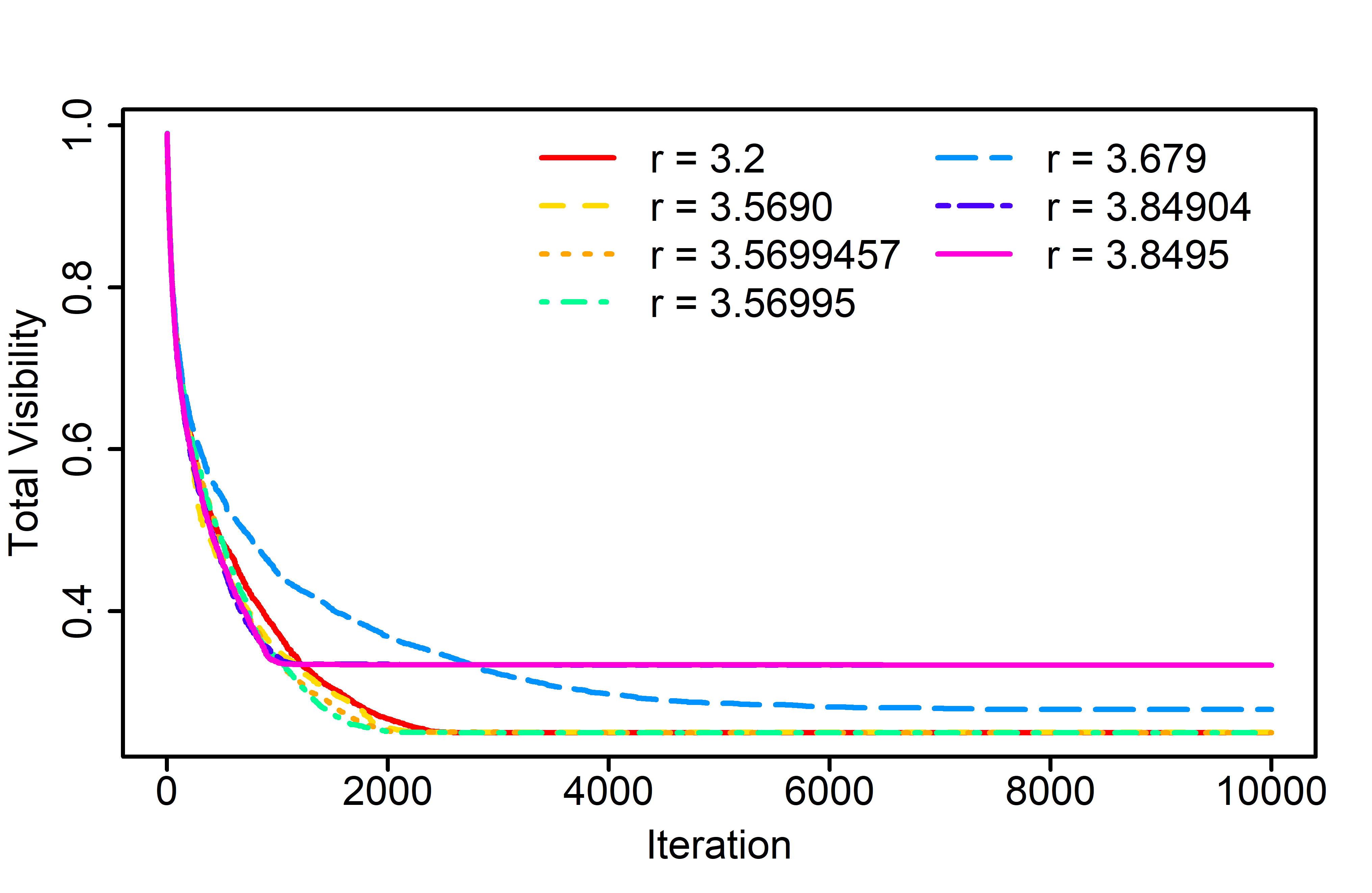}
\caption{The total visibility, the fitness value of the best genomes of the population,  is represented as a function of the computational steps for different values of the growth rate $r$. As it can be observed, the transition time to the corresponding total visibility value depends on $r$, being slower for $r = 3.679$, near the crossing point.}
\label{Figtimestep}
\end{figure}

\end{document}